\documentstyle[12pt]{article}
\def\1{{\chi}}

\begin{document}
%\centerline{Islamic University Journal\hfill Vol. 5, No. 2 June 1997}
%\hrule
%\vskip 2in
\title {{Not each sequential effect algebra is sharply
dominating}\thanks{This project is supported by Natural Science
Found of China (10771191 and 10471124).}}
%\vskip 1.5in
\author {Shen Jun$^{1,2}$, Wu Junde$^{1}$\date{}\thanks{E-mail: wjd@zju.edu.cn}}
\maketitle
$^1${\small\it Department of Mathematics, Zhejiang
University, Hangzhou 310027, P. R. China}

$^2${\small\it Department of Mathematics, Anhui Normal University,
Wuhu 241003, P. R. China}

\begin{abstract} {Let $E$ be an effect algebra and $E_S$ be the set of all sharp elements of $E$. $E$ is said to be sharply dominating if for each $a\in E$ there exists a smallest
element $\widehat{a}\in E_s$ such that $a\leq \widehat{a}$. In 2002,
Professors Gudder and Greechie proved that each $\sigma$-sequential
effect algebra is sharply dominating. In 2005, Professor Gudder
presented 25 open problems in International Journal of Theoretical
Physics, Vol. 44, 2199-2205, the 3th problem asked: Is each
sequential effect algebra sharply dominating? Now, we construct an
example to answer the problem negatively.}
\end{abstract}

\vskip0.1in

Effect algebra is an important model for studying the unsharp
quantum logic (see [1]). In 2001, in order to study quantum
measurement theory, Professor Gudder began to consider the
sequential product of two measurements $A$ and $B$ (see [2]). In
2002, moreover, Professors Gudder and Greechie introduced the
abstract sequential effect algebra structure and studied its some
important properties. In particular, they proved that each
$\sigma$-sequential effect algebra is sharply dominating ([3,
Theorem 6.3]). In 2005, Professor Gudder presented 25 open problems
in [4] to motive the study of sequential effect algebra theory, the
3th problem asked: Is each sequential effect algebra sharply
dominating? Now, we construct an example to answer the problem
negatively.

\vskip 0.1 in

First, we need the following basic definitions and results for
effect algebras and sequential effect algebras.

\vskip 0.1 in

An {\it effect algebra} is a system $(E,0,1, \oplus)$, where 0 and 1
are distinct elements of $E$ and $\oplus$ is a partial binary
operation on $E$ satisfying that [1]:

(EA1) If $a\oplus b$ is defined, then $b\oplus a$ is defined and
$b\oplus a=a\oplus b$.

(EA2) If $a\oplus (b\oplus c)$ is defined, then $(a\oplus b)\oplus
c$ is defined and $$(a\oplus b)\oplus c=a\oplus (b\oplus c).$$

(EA3) For each $a\in E$, there exists a unique element $b\in E$ such
that $a\oplus b=1$.

(EA4) If $a\oplus 1$ is defined, then $a=0$.

\vskip 0.1 in

In an effect algebra $(E,0,1, \oplus)$, if $a\oplus b$ is defined,
we write $a\bot b$. For each $a\in (E,0,1, \oplus)$, it follows from
(EA3) that there exists a unique element $b\in E$ such that $a\oplus
b=1$, we denote $b$ by $a'$. Let $a, b\in (E,0,1, \oplus)$, if there
exists a $c\in E$ such that $a\bot c$ and $a\oplus c=b$, then we say
that $a\leq b$. It follows from [1] that $\leq $ is a partial order
of $(E,0,1, \oplus)$ and satisfies that for each $a\in E$, $0\leq
a\leq 1$, $a\bot b$ if and only if $a\leq b'$.

Let $(E,0,1, \oplus , \circ)$ be an effect algebra and $a\in E$. If
$a\wedge a'=0$, then $a$ is said to be a {\it sharp element} of $E$.
The set $E_S=\{x\in E|\ x\wedge x'=0\}$ is called the set of all
sharp elements of $E$ (see [5-6]). The effect algebra $(E,0,1,
\oplus , \circ)$ is called {\it sharply dominating} if for each
$a\in E$ there exists a smallest sharp element $\widehat{a}\in E_s$
such that $a\leq \widehat{a}$. That is, if $b\in E_s$ satisfies
$a\leq b$, then $\widehat{a}\leq b$. An important example of sharply
dominating effect algebras is the standard Hilbert space effect
algebra $\cal E(H)$ of positive linear operators on a complex
Hilbert space $\cal H$ with norm less than 1([5-6]). The sharply
dominating effect algebras have many nice properties, for example,
recently, Riecanova and Wu showed that sharply dominating
Archimedean atomic lattice effect algebras can be characterized by
the property called basic decomposition of elements, etc (see [5]).

\vskip 0.1 in

A {\it sequential effect algebra} is an effect algebra $(E,0,1,
\oplus)$ and another binary operation $\circ $ defined on $(E,0,1,
\oplus)$ satisfying [3]:

(SEA1) The map $b\mapsto a\circ b$ is additive for each $a\in E$,
that is, if $b\bot c$, then $a\circ b\bot a\circ c$ and $a\circ
(b\oplus c)=a\circ b\oplus a\circ c$.

(SEA2) $1\circ a=a$ for each $a\in E$.

(SEA3) If $a\circ b=0$, then $a\circ b=b\circ a$.

(SEA4) If $a\circ b=b\circ a$, then $a\circ b'=b'\circ a$ and
$a\circ (b\circ c)=(a\circ b)\circ c$ for each $c\in E$.

(SEA5) If $c\circ a=a\circ c$ and $c\circ b=b\circ c$, then
$c\circ(a\circ b)=(a\circ b)\circ c$ and $c\circ(a\oplus b)=(a\oplus
b)\circ c$ whenever $a\bot b$.

\vskip 0.1 in

Let $(E,0,1, \oplus, \circ)$ be a sequential effect algebra. Then
the operation $\circ$ is said to be a sequential product on $(E,0,1,
\oplus, \circ)$. If $a, b\in (E,0,1, \oplus, \circ)$ and $a\circ
b=b\circ a$, then $a$ and $b$ is said to be {\it sequentially
independent} and denoted by $a|b$ (see [2-3]). The sequential effect
algebra is an important and interesting mathematical model for
studying the quantum measurement theory [2-4, 7-8].

Let $(E,0,1, \oplus , \circ)$ be a sequential effect algebra. If
$a\in E$, then it follows from ([3,  Lemma 3.2]) that $a$ is a sharp
element of $(E,0,1, \oplus , \circ)$ iff $a\circ a=a$.

A $\sigma$-{\it effect algebra} is an effect algebra $(E,0,1,
\oplus)$ such that $a_1\geq a_2\geq a_3\cdots$ implies that
$\bigwedge a_i$ exists in $E$. A $\sigma$-{\it sequential effect
algebra} $(E,0,1, \oplus , \circ)$ is a sequential effect algebra
and is a $\sigma$-{\it effect algebra} satisfying [3]:

(1). If $a_1\geq a_2\geq a_3\cdots$, then $b\circ (\bigwedge
a_i)=\bigwedge (b\circ a_i)$ for each $b\in E$;

(2). If $a_1\geq a_2\geq a_3\cdots$ and $b|a_i, i=1, 2, \cdots$,
then $b|(\bigwedge a_i)$.

It is known that ${\cal E(H)}$ is a $\sigma$-sequential effect
algebra (see [3]).

\vskip 0.1 in

In 2002, Professors Gudder and Greechie proved the following
important conclusion ([3, Theorem 6.3]): Every $\sigma$-sequential
effect algebra is sharply dominating.

\vskip 0.1 in

In 2005, by the motivation of the above result, Professor Gudder
asked ([4, Problem 3]): Is each sequential effect algebra sharply
dominating?

\vskip 0.1 in

Now, we construct a sequential effect algebra which is not sharply
dominating, thus, we answer the above problem negatively.

\vskip 0.2 in

Let $E_0=\{0,1,a_n,b_n,c_{\wedge,n},d_{\wedge,n}|\ n\in
{\mathbf{N}}^+,\wedge\in\Lambda\}$, where $\mathbf{N}^+$ be the
positive integer set and $\Lambda$ be the set of all finite nonempty
subsets of $\mathbf{N}^+$. First, we define a partial binary
operation $\oplus$ on $E_0$ as following (when we write $x\oplus
y=z$, we always mean that $x\oplus y=z=y\oplus x$):

For each $x\in E_0$, $0\oplus x=x$,

$a_n\oplus a_m=a_{n+m}$,

For $n<m$, $a_n\oplus b_m=b_{m-n}$, $a_n\oplus b_n=1$,

$a_n\oplus c_{\wedge,m}=c_{\wedge,n+m}$,

For $n<m$, $a_n\oplus d_{\wedge,m}=d_{\wedge,m-n}$,

For $\wedge \cap I=\emptyset$, $c_{\wedge,n}\oplus c_{I,m}=c_{\wedge
\cup I,m+n-1}$,

For $\wedge \subset I\ and\ n\leq m$, $c_{\wedge,n}\oplus
d_{I,m}=d_{I\backslash\wedge , m-n+1} (when\ \wedge \neq I)\ or\
b_{m-n}(when\ \wedge=I\ and\ n<m)\ or\ 1 (when\ \wedge=I\ and\ n=m)
$.

No other $\oplus$ operation is defined.

Next, we define a binary operation  $\circ$  on $E_0$ as following
(when we write $x\circ y=z$, we always mean that $x\circ y=z=y\circ
x$):

For each $x\in E_0$, $0\circ x=0$, $1\circ x=x$,

$a_n\circ a_m=0$, $a_n\circ b_m=a_n$, $b_n\circ b_m=b_{m+n}$,
$a_n\circ c_{\wedge,m}=0$, $c_{\wedge,n}\circ b_m=c_{\wedge,n}$,
$a_n\circ d_{\wedge,m}=a_n$, $b_n\circ d_{\wedge,m}=d_{\wedge,m+n}$,
$d_{\wedge,n}\circ d_{I,m}=d_{\wedge \cup I,n+m-1}$,

$c_{\wedge,n}\circ c_{I,m}=c_{\wedge \cap I,1} (when\ \wedge \cap
I\neq\emptyset)\ or\ 0 (when\ \wedge \cap I=\emptyset)$,

$c_{\wedge,n}\circ d_{I,m}=c_{\wedge\backslash I,n}(when\
\wedge\backslash I\neq\emptyset)\ or\ a_{n-1} (when\
\wedge\backslash I=\emptyset\ and\ n>1)\ or\ 0 (when\
\wedge\backslash I=\emptyset\ and\ n=1)$.

{\bf Proposition 1.} $(E_0,0,1, \oplus , \circ)$ {\it is a
sequential effect algebra}.

{\bf Proof.} First we verify that $(E_0,0,1, \oplus)$ is an effect
algebra.

(EA1) and (EA4) are trivial.

We verify (EA2), we omit the trivial cases about 0,1:

$a_n\oplus (a_m\oplus a_{k})=(a_n\oplus a_m)\oplus a_{k}=a_{k+m+n}$.

$a_n\oplus (a_m\oplus c_{\wedge,k})=(a_n\oplus a_m)\oplus
c_{\wedge,k}=c_{\wedge,k+m+n}$.

Each $a_n\oplus (a_m\oplus b_{k})$ or $(a_n\oplus a_m)\oplus b_{k}$
is defined iff $n+m\leq k$, $a_n\oplus (a_m\oplus b_{k})=(a_n\oplus
a_m)\oplus b_{k}=b_{k-m-n}(when\ m+n<k)\ or\ 1(when\ m+n=k)$.

Each $a_n\oplus (a_m\oplus d_{\wedge,k})$ or $(a_n\oplus a_m)\oplus
d_{\wedge,k}$ is defined iff $n+m< k$, $a_n\oplus (a_m\oplus
d_{\wedge,k})=(a_n\oplus a_m)\oplus d_{\wedge,k}=d_{\wedge,k-m-n}$.

Each $a_n\oplus (c_{\wedge,m}\oplus d_{I,k})$ or $(a_n\oplus
c_{\wedge,m})\oplus d_{I,k}$ or $(a_n\oplus d_{I,k})\oplus
c_{\wedge,m}$ is defined iff $\wedge \subset I\ and\ n+m\leq k$,
$a_n\oplus (c_{\wedge,m}\oplus d_{I,k})=(a_n\oplus
c_{\wedge,m})\oplus d_{I,k}=(a_n\oplus d_{I,k})\oplus
c_{\wedge,m}=d_{I\backslash\wedge,k-m-n+1}(when\ \wedge \neq I)\ or\
b_{k-m-n}(when\ \wedge=I\ and\ m+n<k)\ or\ 1(when\ \wedge=I\
 and\ m+n=k)$.

Each $a_n\oplus (c_{\wedge,m}\oplus c_{I,k})$ or $(a_n\oplus
c_{\wedge,m})\oplus c_{I,k}$ is defined iff $\wedge \cap
I=\emptyset$, $a_n\oplus (c_{\wedge,m}\oplus c_{I,k})=(a_n\oplus
c_{\wedge,m})\oplus c_{I,k}=c_{\wedge \cup I,n+m+k-1}$.

Each $c_{\wedge,n}\oplus (c_{I,m}\oplus c_{Y,k})$ or
$(c_{\wedge,n}\oplus c_{Y,k})\oplus c_{I,m}$ is defined iff
$\wedge\cap I\ and\ \wedge\cap Y\ and\ Y\cap I\ are\ all\
\emptyset$, $c_{\wedge,n}\oplus (c_{I,m}\oplus
c_{Y,k})=(c_{\wedge,n}\oplus c_{Y,k})\oplus c_{I,m}=c_{\wedge\cup
I\cup Y,n+m+k-2}$.

Each $c_{\wedge,n}\oplus (c_{I,m}\oplus d_{Y,k})$ or
$(c_{\wedge,n}\oplus c_{I,m})\oplus d_{Y,k}$ is defined iff
$\wedge\cap I=\emptyset\ and\ \wedge\cup I\subset Y\ and\ n+m\leq
k+1$, $c_{\wedge,n}\oplus (c_{I,m}\oplus
d_{Y,k})=(c_{\wedge,n}\oplus c_{I,m})\oplus
d_{Y,k}=d_{Y\backslash(\wedge\cup I),k-m-n+2}(when\ \wedge\cup I\neq
Y)\ or\ b_{k-n-m+1}(when\ \wedge\cup I=Y\ and\ m+n<k+1)\ or\ 1(when\
\wedge\cup I=Y\ and\ m+n=k+1)$.

Thus, (EA2) is proved. We verify (EA3):

$a_n\oplus b_n=1$, $c_{\wedge,n}\oplus d_{\wedge,n}=1$.

So $(E,0,1, \oplus)$ is an effect algebra.

We now verify that $(E,0,1, \oplus , \circ)$ is a sequential effect
algebra.

(SEA2) and (SEA3) and (SEA5) are trivial.

We verify (SEA1), we omit the trivial cases about 0,1:

$a_n\circ (a_m\oplus a_{k})=a_n\circ a_m\oplus a_n\circ a_{k}=0$,

$b_n\circ (a_m\oplus a_{k})=b_n\circ a_m\oplus b_n\circ
a_{k}=a_{m+k}$,

$c_{\wedge,n}\circ (a_m\oplus a_{k})=c_{\wedge,n}\circ a_m\oplus
c_{\wedge,n}\circ a_{k}=0$,

$d_{\wedge,n}\circ (a_m\oplus a_{k})=d_{\wedge,n}\circ a_m\oplus
d_{\wedge,n}\circ a_{k}=a_{m+k}$.

$a_n\circ (a_m\oplus c_{\wedge,k})=a_n\circ a_m\oplus a_n\circ
c_{\wedge,k}=0$,

$b_n\circ (a_m\oplus c_{\wedge,k})=b_n\circ a_m\oplus b_n\circ
c_{\wedge,k}=c_{\wedge,m+k}$,

$c_{I,n}\circ (a_m\oplus c_{\wedge,k})=c_{I,n}\circ a_m\oplus
c_{I,n}\circ c_{\wedge,k}=c_{\wedge \cap I,1}(when\ \wedge \cap
I\neq\emptyset)\ or\ 0(when\ \wedge \cap I=\emptyset)$,

$d_{I,n}\circ (a_m\oplus c_{\wedge,k})=d_{I,n}\circ a_m\oplus
d_{I,n}\circ c_{\wedge,k}=c_{\wedge\backslash I,m+k}(when\
\wedge\backslash I\neq\emptyset)\ or\ a_{m+k-1}(when\
\wedge\backslash I=\emptyset)$.

For $m<k$,

$a_n\circ (a_m\oplus d_{\wedge,k})=a_n\circ a_m\oplus a_n\circ
d_{\wedge,k}=a_n$,

$b_n\circ (a_m\oplus d_{\wedge,k})=b_n\circ a_m\oplus b_n\circ
d_{\wedge,k}=d_{\wedge,n+k-m}$,

$c_{I,n}\circ (a_m\oplus d_{\wedge,k})=c_{I,n}\circ a_m\oplus
c_{I,n}\circ d_{\wedge,k}=c_{I\backslash\wedge,n}(when\
I\backslash\wedge\neq\emptyset)\ or\ a_{n-1}(when\
I\backslash\wedge=\emptyset\ and\ n>1)\ or\ 0(when\
I\backslash\wedge=\emptyset\ and\ n=1)$,

$d_{I,n}\circ (a_m\oplus d_{\wedge,k})=d_{I,n}\circ a_m\oplus
d_{I,n}\circ d_{\wedge,k}=d_{\wedge\cup I,n+k-m-1}$.

For $m\leq k$,

$a_n\circ (a_m\oplus b_{k})=a_n\circ a_m\oplus a_n\circ b_{k}=a_n$,

$b_n\circ (a_m\oplus b_{k})=b_n\circ a_m\oplus b_n\circ
b_{k}=b_{n+k-m}$,

$c_{\wedge,n}\circ (a_m\oplus b_{k})=c_{\wedge,n}\circ a_m\oplus
c_{\wedge,n}\circ b_{k}=c_{\wedge,n}$,

$d_{\wedge,n}\circ (a_m\oplus b_{k})=d_{\wedge,n}\circ a_m\oplus
d_{\wedge,n}\circ b_{k}=d_{\wedge,n+k-m}$.

For $\wedge \cap I=\emptyset$,

$a_n\circ (c_{\wedge,m}\oplus c_{I,k})=a_n\circ c_{\wedge,m}\oplus
a_n\circ c_{I,k}=0$,

$b_n\circ (c_{\wedge,m}\oplus c_{I,k})=b_n\circ c_{\wedge,m}\oplus
b_n\circ c_{I,k}=c_{\wedge\cup I,m+k-1}$,

$c_{Y,n}\circ (c_{\wedge,m}\oplus c_{I,k})=c_{Y,n}\circ
c_{\wedge,m}\oplus c_{Y,n}\circ c_{I,k}=c_{Y\cap(\wedge\cup
I),1}(when\ Y\cap(\wedge\cup I)\neq\emptyset)\ or\ 0(when\
Y\cap(\wedge\cup I)=\emptyset)$,

$d_{Y,n}\circ (c_{\wedge,m}\oplus c_{I,k})=d_{Y,n}\circ
c_{\wedge,m}\oplus d_{Y,n}\circ c_{I,k}=c_{(\wedge\cup I)\backslash
Y,m+k-1}(when\ (\wedge\cup I)\backslash Y\neq\emptyset)\ or\
a_{m+k-2}(when\ (\wedge\cup I)\backslash Y=\emptyset\ and\ m+k>2)\
or\ 0(when\ (\wedge\cup I)\backslash Y=\emptyset\ and\ m+k=2)$.

For $\wedge \subset I\ and\ m\leq k$,

$a_n\circ (c_{\wedge,m}\oplus d_{I,k})=a_n\circ c_{\wedge,m}\oplus
a_n\circ d_{I,k}=a_n$,

$b_n\circ (c_{\wedge,m}\oplus d_{I,k})=b_n\circ c_{\wedge,m}\oplus
b_n\circ d_{I,k}=d_{I\backslash\wedge,n+k-m+1}(when\ \wedge\neq I)\
or\ b_{n+k-m}(when\ \wedge=I)$,

$c_{Y,n}\circ (c_{\wedge,m}\oplus d_{I,k})=c_{Y,n}\circ
c_{\wedge,m}\oplus c_{Y,n}\circ
d_{I,k}=c_{Y\backslash(I\backslash\wedge), n}(when\
Y\backslash(I\backslash\wedge)\neq\emptyset)\ or\ a_{n-1}(when\
Y\backslash(I\backslash\wedge)=\emptyset\ and\ n>1)\ or\ 0(when\
Y\backslash(I\backslash\wedge)=\emptyset\ and\ n=1)$,

$d_{Y,n}\circ (c_{\wedge,m}\oplus d_{I,k})=d_{Y,n}\circ
c_{\wedge,m}\oplus d_{Y,n}\circ
d_{I,k}=d_{Y\cup(I\backslash\wedge),n+k-m}$.

Thus, (SEA1) is proved. We verify (SEA4), we omit the trivial cases
about 0,1:

$a_n\circ (a_m\circ a_{k})=(a_n\circ a_m)\circ a_{k}=0$,

$a_n\circ (a_m\circ b_{k})=b_{k}\circ (a_n\circ a_m)=a_m\circ
(a_n\circ b_{k})=0$,

$a_n\circ (a_m\circ c_{\wedge,k})=c_{\wedge,k}\circ (a_n\circ
a_m)=a_m\circ (a_n\circ c_{\wedge,k})=0$,

$a_n\circ (a_m\circ d_{\wedge,k})=d_{\wedge,k}\circ (a_n\circ
a_m)=a_m\circ (a_n\circ d_{\wedge,k})=0$,

$a_n\circ (b_m\circ b_{k})=b_{k}\circ (a_n\circ b_m)=b_m\circ
(a_n\circ b_{k})=a_n$,

$a_n\circ (b_m\circ c_{\wedge,k})=c_{\wedge,k}\circ (a_n\circ
b_m)=b_m\circ (a_n\circ c_{\wedge,k})=0$,

$a_n\circ (b_m\circ d_{\wedge,k})=d_{\wedge,k}\circ (a_n\circ
b_m)=b_m\circ (a_n\circ d_{\wedge,k})=a_n$,

$a_n\circ (c_{I,m}\circ c_{\wedge,k})=c_{\wedge,k}\circ (a_n\circ
c_{I,m})=c_{I,m}\circ (a_n\circ c_{\wedge,k})=0$,

$a_n\circ (c_{I,m}\circ d_{\wedge,k})=d_{\wedge,k}\circ (a_n\circ
c_{I,m})=c_{I,m}\circ (a_n\circ d_{\wedge,k})=0$,

$a_n\circ (d_{I,m}\circ d_{\wedge,k})=d_{\wedge,k}\circ (a_n\circ
d_{I,m})=d_{I,m}\circ (a_n\circ d_{\wedge,k})=a_n$,

$b_n\circ (b_m\circ b_{k})=b_{k}\circ (b_n\circ b_m)=b_{m+n+k}$,

$b_n\circ (b_m\circ c_{\wedge,k})=c_{\wedge,k}\circ (b_n\circ
b_m)=b_m\circ (b_n\circ c_{\wedge,k})=c_{\wedge,k}$,

$b_n\circ (b_m\circ d_{\wedge,k})=d_{\wedge,k}\circ (b_n\circ
b_m)=b_m\circ (b_n\circ d_{\wedge,k})=d_{\wedge,n+m+k}$,

$b_n\circ (c_{I,m}\circ c_{\wedge,k})=c_{\wedge,k}\circ (b_n\circ
c_{I,m})=c_{I,m}\circ (b_n\circ
c_{\wedge,k})=c_{I\cap\wedge,1}(when\ I\cap\wedge\neq\emptyset)\ or\
0(when\ I\cap\wedge=\emptyset)$,

$b_n\circ (c_{I,m}\circ d_{\wedge,k})=d_{\wedge,k}\circ (b_n\circ
c_{I,m})=c_{I,m}\circ (b_n\circ
d_{\wedge,k})=c_{I\backslash\wedge,m}(when\
I\backslash\wedge\neq\emptyset)\ or\ a_{m-1}(when\
I\backslash\wedge=\emptyset\ and\ m>1)\ or\ 0(when\
I\backslash\wedge=\emptyset\ and\ m=1)$,

$b_n\circ (d_{I,m}\circ d_{\wedge,k})=d_{\wedge,k}\circ (b_n\circ
d_{I,m})=d_{I,m}\circ (b_n\circ
d_{\wedge,k})=d_{I\cup\wedge,n+m+k-1}$,

$c_{Y,n}\circ (c_{I,m}\circ c_{\wedge,k})=c_{\wedge,k}\circ
(c_{Y,n}\circ c_{I,m})=c_{Y\cap I\cap\wedge,1}(when\ Y\cap
I\cap\wedge\neq\emptyset)\ or\ 0(when\ Y\cap
I\cap\wedge=\emptyset)$,

$c_{Y,n}\circ (c_{I,m}\circ d_{\wedge,k})=d_{\wedge,k}\circ
(c_{Y,n}\circ c_{I,m})=c_{I,m}\circ (c_{Y,n}\circ
d_{\wedge,k})=c_{(Y\cap I)\backslash\wedge,1}(when\ (Y\cap
I)\backslash\wedge\neq\emptyset)\ or\ 0(when\ (Y\cap
I)\backslash\wedge=\emptyset)$,

$c_{Y,n}\circ (d_{I,m}\circ d_{\wedge,k})=d_{\wedge,k}\circ
(c_{Y,n}\circ d_{I,m})=d_{I,m}\circ (c_{Y,n}\circ
d_{\wedge,k})=c_{Y\backslash(\wedge\cup I),n}(when\
Y\backslash(\wedge\cup I)\neq\emptyset)\ or\ a_{n-1}(when\
Y\backslash(\wedge\cup I)=\emptyset\ and\ n>1)\ or\ 0(when\
Y\backslash(\wedge\cup I)=\emptyset\ and\ n=1)$,

$d_{Y,n}\circ (d_{I,m}\circ d_{\wedge,k})=d_{\wedge,k}\circ
(d_{Y,n}\circ d_{I,m})=d_{\wedge\cup I\cup Y,n+m+k-2}$.

(SEA4) is proved and so $(E_0,0,1, \oplus , \circ)$ is a sequential
effect algebra.

Our main result is:

{\bf Theorem 1.} Not each sequential effect algebra is sharply
dominating.

{\bf Proof.} In fact, in the sequential effect algebra $(E_0,0,1,
\oplus , \circ)$, its all sharp elements is the set
$E_s=\{0,1,c_{\wedge,1},d_{\wedge,1}|\ \wedge\in\Lambda$, where
$\Lambda$ is the set of all finite nonempty subsets of $\mathbf{N}^+
\}$. Note that when $\wedge_1\subset\wedge_2$ and
$\wedge_1\neq\wedge_2$, $c_{\wedge_1,1}\oplus
c_{\wedge_2\backslash\wedge_1,1}=c_{\wedge_2,1}$,
$d_{\wedge_2,1}\oplus
c_{\wedge_2\backslash\wedge_1,1}=d_{\wedge_1,1}$, so
$c_{\wedge_1,1}<c_{\wedge_2,1}$, $d_{\wedge_2,1}<d_{\wedge_1,1}$.
For each finite subset $\wedge$ of ${\mathbf{N}}^+$, $a_1\oplus
d_{\wedge,2}=d_{\wedge,1}$, so $a_1<d_{\wedge,1}$, and there is no
comparison relation between $a_1$ and $c_{\wedge,1}$. So the set of
elements in $E_s$ larger than $a_1$ is $A=\{1,d_{\wedge,1}|\
\wedge\in \Lambda\}$, nevertheless, there is no smallest element in
$A$. Thus, $(E_0,0,1, \oplus , \circ)$ is not sharply dominating and
the theorem is proved.

\vskip 0.1 in

Moreover, we show that the sequential effect algebra $(E_0, 0, 1,
\oplus, \circ)$ in Proposition 1 is not even a $\sigma$-effect
algebra. At first, we need the following theorem:

\vskip 0.1 in

{\bf Theorem 2.} Let $(E,0,1, \oplus , \circ)$ be a sequential
effect algebra, $I$ be an index set, $\{a_\alpha\}_{\alpha\in
I}\subset E_s$.

(1) If $\bigwedge\limits_{\alpha\in I}a_\alpha$ exists, then
$\bigwedge\limits_{\alpha\in I}a_\alpha\in E_s$;

(2) if $\bigvee\limits_{\alpha\in I}a_\alpha$ exists, then
$\bigvee\limits_{\alpha\in I}a_\alpha\in E_s$.

{\bf Proof.} Just the same as the proof of [3] corollary 4.3.

\vskip 0.1 in

{\bf Proposition 2.} $(E_0,0,1, \oplus)$ is not a $\sigma$-effect
algebra.

{\bf Proof.} Let $\{\wedge_i\}_{i\in {\mathbf{N}}^+}$ be a strictly
increasing sequence of finite nonempty subsets of $\mathbf{N}^+$. We
note from the proof of Theorem 1 that $$\hbox{$\{d_{\wedge_i,1}|\
i\in {\mathbf{N}}^+\}\subset E_s$ and satisfying
$d_{\wedge_1,1}>d_{\wedge_2,1}>\cdots>d_{\wedge_n,1}>\cdots\ .$}$$

If $(E_0,0,1, \oplus)$ is a $\sigma$-effect algebra, then
$\bigwedge\limits_{i\in {\mathbf{N}}^+}d_{\wedge_i,1}$ will exist,
and it follows from Theorem 2 that $\bigwedge\limits_{i\in
{\mathbf{N}}^+}d_{\wedge_i,1}\in E_s$.

By the proof of Theorem 1 again, we have $a_1<d_{\wedge,1}$, so
$a_1\leq\bigwedge\limits_{i\in {\mathbf{N}}^+}d_{\wedge_i,1}$. Note
that there is no comparison relation between $a_1$ and
$c_{\wedge,1}$ (proof of Theorem 1), so $\bigwedge\limits_{i\in
{\mathbf{N}}^+}d_{\wedge_i,1}$ is not $c_{\wedge,1}$. Also, it is
obvious that $\bigwedge\limits_{i\in {\mathbf{N}}^+}d_{\wedge_i,1}$
is not 0 or 1.

It follows from above and $E_s=\{0,1,c_{\wedge,1},d_{\wedge,1}|\
\wedge\in\Lambda,$ where $\Lambda$ is the set of all finite nonempty
subsets of $\mathbf{N}^+ \}$ that there exists some $\wedge_0$ such
that $d_{\wedge_0,1}=\bigwedge\limits_{i\in
{\mathbf{N}}^+}d_{\wedge_i,1}$. But then we will have
$d_{\wedge_0,1}\leq d_{\wedge_i,1}$ and $\wedge_0\supset\wedge_i$
for all $i\in {\mathbf{N}}^+$, which is impossible since $\wedge_0$
is a finite subset of $\mathbf{N}^+$.

\vskip 0.2 in

\centerline {\bf Acknowledgement}

\vskip 0.2 in

The authors wish to express their thanks to the referee for his
valuable comments and suggestions.

\vskip 0.2 in

\vskip 0.2 in

\centerline{\bf References}

\vskip 0.2 in

\noindent [1]. Foulis, D J, Bennett, M K. Effect algebras and
unsharp quantum logics. Found Phys 24 (1994), 1331-1352.

\noindent [2]. Gudder, S, Nagy, G. Sequential quantum measurements.
J. Math. Phys. 42(2001), 5212-5222.

\noindent [3]. Gudder, S, Greechie, R. Sequential products on effect
algebras. Rep. Math. Phys.  49(2002), 87-111.

\noindent [4]. Gudder, S. Open problems for sequential effect
algebras. Inter. J. Theory. Phys. 44 (2005), 2219-2230.

\noindent [5]. Gudder, S. Sharply dominating effect algebras. Tatra
Mt. Math. Publ., 15(1998), 23-30.

\noindent [6] Riecanova, Z, Wu Junde. States on sharply dominating
effect algebras. Science in China A: Mathematics, 51(2008), 907-914.

\noindent [7]. Gheondea, A, Gudder, S. Sequential product of quantum
effects. Proc. Amer. Math. Soc. 132 (2004), 503-512.

\noindent [8]. Gudder, S, Latr¨¦moli¨¨re, F. Characterization of the
sequential product on quantum effects. J. Math. Phys. 49 (2008),
052106-052112.

\end{document}